\documentclass[runningheads]{llncs}
\usepackage[T1]{fontenc}
\usepackage{graphicx}
\usepackage{enumitem}
\usepackage{amssymb}
\usepackage{algorithmic}
\usepackage{algorithm}
\usepackage{multirow} 

\usepackage{cite}  
\usepackage{hyperref}
\usepackage{bbding}

\usepackage{amsmath}

\begin{document}
\title{SPGL: Enhancing Session-based Recommendation with Single Positive Graph Learning}
\titlerunning{SPGL}
%
\author{Tiantian Liang \and
Zhe Yang \Envelope }
\authorrunning{T. Liang et al.}
%
\institute{School of Computer Science and Technology, Soochow University, Suzhou, China\\ 
\email{yangzhe@suda.edu.cn}}

\maketitle              
\begin{abstract}
Session-based recommendation seeks to forecast the next item a user will be interested in, based on their interaction sequences. Due to limited interaction data, session-based recommendation faces the challenge of limited data availability. Traditional methods enhance feature learning by constructing complex models to generate positive and negative samples. This paper proposes a session-based recommendation model using Single Positive optimization loss and Graph Learning (SPGL) to deal with the problem of data sparsity, high model complexity and weak transferability. SPGL utilizes graph convolutional networks to generate global item representations and batch session representations, effectively capturing intrinsic relationships between items. The use of single positive optimization loss improves uniformity of item representations, thereby enhancing recommendation accuracy. In the intent extractor, SPGL considers the hop count of the adjacency matrix when constructing the directed global graph to fully integrate spatial information. It also takes into account the reverse positional information of items when constructing session representations to incorporate temporal information. Comparative experiments across three benchmark datasets, Tmall, RetailRocket and Diginetica, demonstrate the model's effectiveness.

\end{abstract}
\section{Introduction}
Recommendation systems are crucial for assisting users in discovering relevant information. Traditional methods usually depend on user's long-term historical behavior and personal profile data. However, these methods often underperform in many real-world scenarios where users may not log in or their privacy is protected, resulting in limited available information. Session-based recommendation aims to address this challenge by forecasting the next item of interest for a user based on the current anonymous ordered session sequence.\cite{wang2021survey}. \par
In early session-based recommendation research, Markov chains are introduced to capture temporal information \cite{rendle2010factorizing}. Subsequently, deep learning demonstrates overwhelming advantages in sequential data modeling. Hidasi et al. \cite{hidasi2015session} propose using a multi-layer Gated Recurrent Unit (GRU) to get short-term preferences. Li et al. \cite{li2017neural} introduce NARM, which captures users' sequential patterns and main intentions by employing both global and local recurrent neural networks. Liu et al. \cite{liu2018stamp} introduce STAMP, a model that uses a straightforward MLP network enhanced with an attention mechanism to discern both the overarching and immediate preference of users. However, these methods ignore complex transition relationships within entire session sequences.\par
Graph Neural Network (GNN) is good at grasping intricate item relationships by aggregating information from surrounding nodes. The first GNN-based model, SR-GNN \cite{wu2019session}, constructs item representations using a GNN and combines an attention mechanism to obtain user preference vectors. GCE-GNN \cite{wang2020global} constructs both session graphs and global graphs, leveraging the global graph to capture information pertinent to the ongoing session.\par
Contrastive learning-based methods also show promise in session-based recommendation. For instance, COTREC \cite{xia2021selfco} constructs two graph encoders that supervise each other through contrastive learning, thereby forming more accurate representation and achieving better performance. However, this approach is heavily reliant on the specific model architecture. \par
We propose a session-based recommendation model that employs single positive optimization loss as a form of self-contrastive learning combined with graph convolutional networks. The integration of single positive optimization loss with a general graph model leads to significant improvements in performance. The source code can be accessed on \href{https://github.com/liang-tian-tian/SPGL}{Github}. \par
The key advancements presented in this paper include the following:
\begin{itemize}[nosep]
\item[-] We establish a directed global item graph with a three-hop range. Within a session, the closer the items are to each other, the greater the edge weights. In the global graph, we use the cumulative weights of the edges between the same items across all sessions, thereby enhancing the spatial association between related items. 
\item[-] We design an intent extractor that includes attention mechanisms and graph convolutional networks to fully capture item information. By incorporating reverse positional information, we can more accurately capture session representations to better align with user intentions.
\item[-] We introduce a single positive optimization loss. Each item is treated as a positive sample, with the remainder of the items classified as negative samples. This approach improves model performance and simplifies the modeling process. 
\item[-] Through experimentation on benchmark datasets, SPGL demonstrates outstanding performance, validating its effectiveness for session-based recommendation.
\end{itemize}

\section{Related Work}
Predicting the next possible item based on a user's recent interactions has become a popular research topic. Graph neural network and contrastive learning have shown significant advantages in session recommendation and are closely connected to the model presented in this paper.
\subsection{Graph Neural Network}
Recently, research has largely concentrated on utilizing GNN to extract relationships within sessions. Research indicates that GNN-based methods achieve better performance than sequence-based models. SR-GNN\cite{wu2019session} captures embeddings on the item graph through a gated graph neural network and uses an attention mechanism to predict the next item. Zhou and Pan et al. \cite{zhou2021temporal} develop dynamic graphs using timestamps to model evolving preferences. GCE-GNN \cite{wang2020global} enhances session embeddings by integrating convolutional techniques specific to both session graphs and global neighbor graphs. G\textsuperscript{3}-SR\cite{deng2022g} extracts global inter-item relationships through unsupervised pre-training. SPARE \cite{peintner2023spare} adopts a new strategy of constructing a global item graph using the shortest path. HICN \cite{sun2024exploiting} designs a sequential hypergraph and employs a hypergraph convolution network to integrate different types of information within the graph. \par
This paper leverages graph convolutional networks to capture global information and combines attention mechanisms with reverse positional information to enhance session representations.
\subsection{Contrastive Learning}
Contrastive learning seeks to align similar samples while distancing unrelated ones, thereby yielding robust feature representations. The self-supervised approach has gained extensive adoption across recommendation systems. GCC \cite{qiu2020gcc} proposes subgraph instance discrimination and utilizes contrastive learning to develop structural representations that are both inherent and portable. Yao et al. \cite{yao2021self} propose a multi-task contrastive learning method designed for dual-tower models. S\textsuperscript{2}-DHCN\cite{xia2021self} designs a contrastive learning mechanism through another graph convolutional neural network model to enhance hypergraph modeling. COTREC\cite{xia2021selfco} extends session-based graphs through contrastive learning to demonstrate internal and external connectivity of sessions. CORE \cite{hou2022core} uses a contrastive representation alignment strategy to unify the representation space of session embeddings.\par
In the image processing domain, self-contrastive learning \cite{bae2023self} processes a single instance (e.g., an image) to generate multiple augmented views, treating these views as positive samples since they come from the same instance, while features from distinct instances are considered as negative samples. Self-contrastive learning is a specialized form of contrastive learning that leverages augmented views of a single sample for self-comparison, thus eliminating the need for external labels. The paper simplifies self-contrastive learning by applying a single positive optimization loss, where the positive sample is itself and all other items are treated as negative samples. This approach reduces computational costs by eliminating the need for generating and processing multiple augmented views, while promoting a uniform distribution of item representations.
\section{Method}
This section outlines the problem statement in session-based recommendation, presents an overview of the SPGL model, and details its key components.
\subsection{Problem Statement}
In session-based recommendation, all candidate items are denoted as a set $V=\{v_{1}, v_{2},..., v_{n}\}$, with $n$ indicating the total quantity of candidate items. $s$ represents a list of $m$ items arranged in sequential order, defined as $s=\{v_{1}^s, v_{2}^s, ..., v_{m}^s\}$, where $v_{k}^s \in V$ represents the item clicked by the user at the $k$-th position in the current session $s$, where $k$ ranges from 1 to $m$. The embedding representation for all items is indicated as $X_{V}^{(l)} \in \mathbb{R}^{n \times d^{(l)}}$, with the initial representation $X_{V}^{(0)}$ being randomly assigned from a uniform distribution. The goal of session-based recommendation is to forecast the subsequent item $v_{m+1}^s$ from the candidate items $V$ based on the sequence of the current session $s$. The prediction layer of the session-based recommendation model scores and ranks the candidate items for each session, denoted as $R=\{ r_{1}^s, r_{2}^s, ..., r_{n}^s\}$, where $r_{i}^s$ denotes the $i$-th candidate item for session $s$. Subsequently, the top $k$ candidate items form the recommendation list.
\subsection{Overview of SPGL}
\begin{figure}
\includegraphics[width=\textwidth]{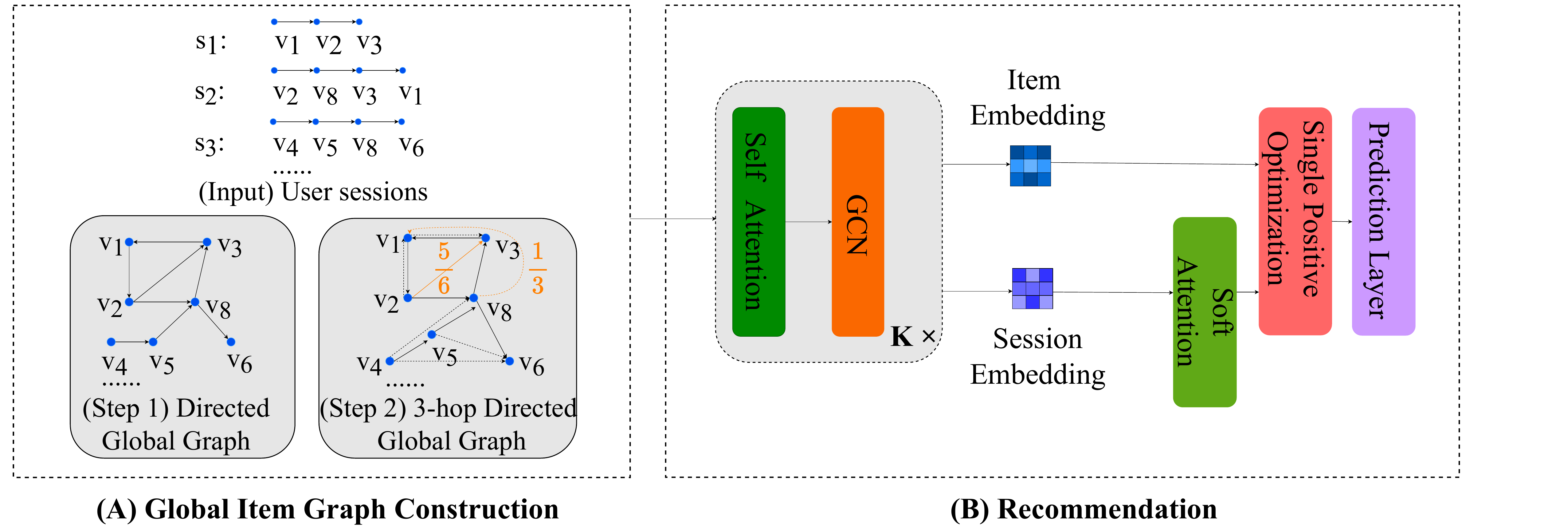}
\caption{The design of the SPGL framework. (A) A global directed graph with 3-hop neighbors is formulated based on user interactions. (B) Session representation learning and item embeddings are derived using graph convolutional networks and attention mechanisms. Single positive optimization enhances item representation, and the prediction layer computes relevance scores for candidate items. } \label{fig1}
\end{figure}

The model SPGL's overall workflow is shown in Fig~\ref{fig1}. In Fig.~\ref{fig1}(A), we first construct 1-hop directed connections, represented by solid black lines. Next, we link nodes that are within three steps of each other in user sessions using dashed black lines. The yellow lines highlight specific weights, showing the closer the two items are, the greater the weight. In Fig.~\ref{fig1}(B), item representations are extracted using self-attention mechanisms and graph convolutional networks. By integrating item representations with reverse positional information through a soft attention mechanism, the session representation is obtained. Relevance scores for candidate items are then derived through the prediction layer. During model optimization, the single positive optimization loss supplements the cross-entropy loss,  enhancing the model's ability to effectively capture item characteristics.

\subsection{Global Item Graph}
The global item graph aims to capture transitions to derive item embeddings across all sessions. Specifically, the global item graph is  formulated  from the $\epsilon$-neighborhood collection of items across all sessions. Given a set of sessions $U=\{s_{1}, s_{2}, ..., s_{l}\}$, where $l$ denotes the overall count of sessions, the $\epsilon$-neighborhood set $N_{\epsilon}(v_{i}^p)$ for an item $v_{i}^p$ in session $s_{p}$ is defined as follows:
\begin{equation}
N_{\epsilon}(v_{i}^p) = \{ v_{j}^q | j\in [k-\epsilon, k+\epsilon], \forall v_{k}^q=v_{i}^p, v_{i}^p\in s_{p}, v_{k}^q \in s_{q}, s_{p} \in U, s_{q} \in U\}
\end{equation}
where $k$ represents the position of the item $v_{k}^q$ the user interacted with in session $q$, and $\epsilon$ is utilized to determine the extent of the neighborhood. The global item graph is designed as $G_{g} = (V_{g}, E_{g})$, with $V_{g}$ representing all items in all sessions. For every item $v_{i}$ in $V$, the set of global-item transitions, denoted as $E_{g}$, is specified as: 
\begin{equation}
E_{g}=\{e_{ij}^g|(v_{i},v_{j})|v_{i}\in V,v_{j}\in N_{\epsilon}(v_{i})\}
\end{equation}
Here, $E_{g}$ represents the collection of edges, with each edge associated with a pair of items that appear across all sessions. The weight of each edge in $E_{g}$ is determined by summing the weights from $v_{i}$ to $v_{j}$ across all sessions. The weight from $v_{i}^s$ to $v_{j}^s$ in session $s$ is defined as:
\begin{equation}
Weight(v_{i}^s, v_{j}^s) = \frac{1}{1 + number\,of\,directed \, edges(v_{i}^s \rightarrow v_{j}^s)}
\end{equation}
Typically, items closer to the target item exhibit higher similarity, so the weights between these closer items should be increased. As the distance between two nodes on the graph increases, their relationship weakens. The weight definition effectively highlights similar nodes in the graph. \par
\begin{figure}
\vspace{-0.4cm}
\includegraphics[width=\textwidth]{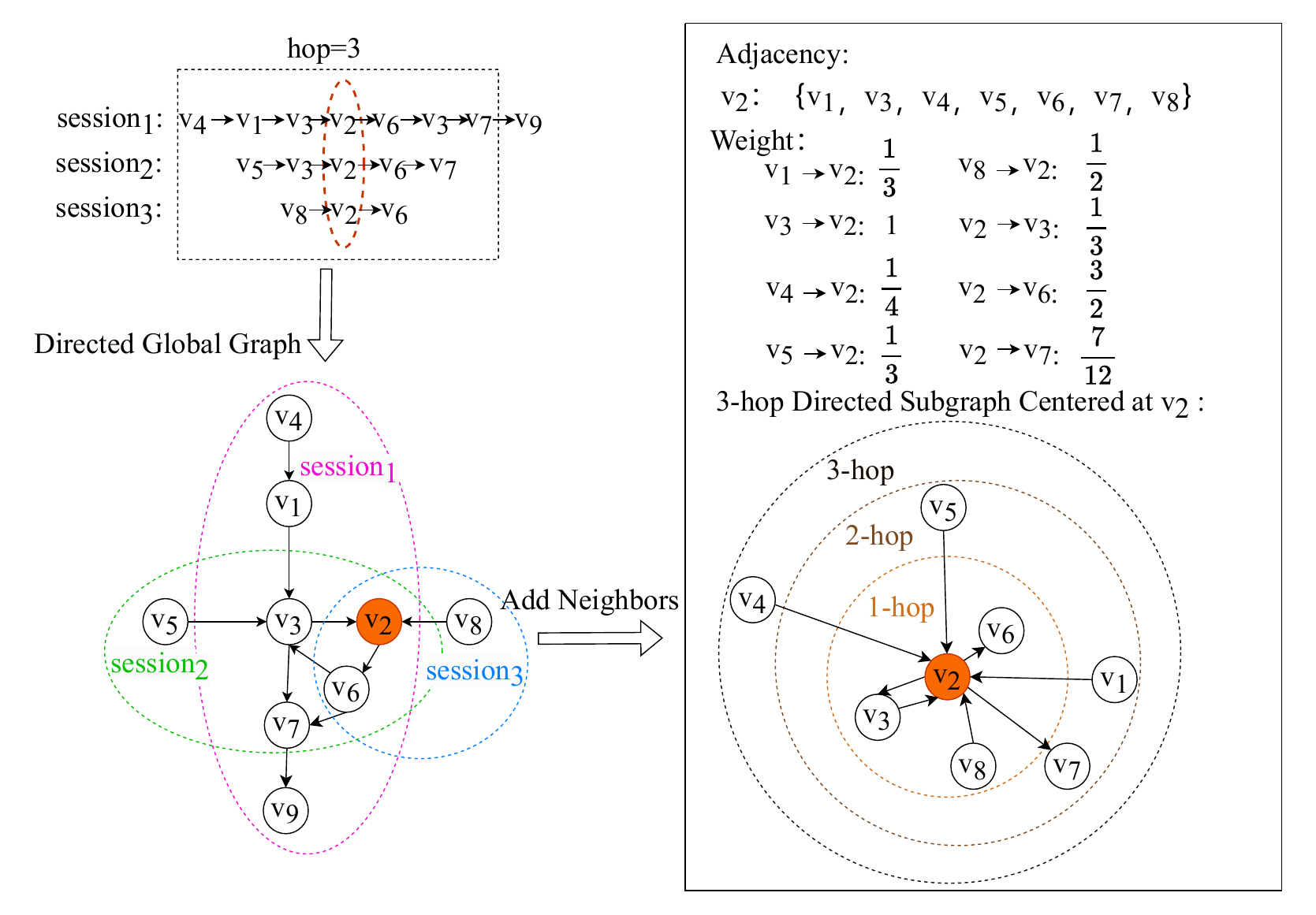}
\caption{Construction of the 3-hop directed global item graph. We first construct the directed global graph across all sessions. Subsequently, we add $\epsilon$-neighbors weighted edges to more precisely capture information. The example illustrates the adjacency and weight calculations for a 3-hop directed subgraph centered at $v_{2}$.}\label{fig2}
\vspace{-0.4cm}
\end{figure}

Figure~\ref{fig2} illustrates an instance of building a global item graph with a neighborhood threshold of $\epsilon=3$. For instance, the total weight of the edge from $v_{3}$ to $v_{2}$ is $\frac{1}{2}+\frac{1}{2}=1$. This is because the edge from $v_{3}$ to $v_{2}$ is directly adjacent in both sessions, leading to a higher weight. $G_{g}$ is a directed global graph within a 3-hop range, capturing the relationships between item pairs while considering temporal information to model user browsing behavior.

\subsection{Intent Extractor}
The Intent extractor is composed of a simplified attention layer coupled with a graph convolution layer. First, the global item graph generates item embeddings that serve as the input to the self-attention layer. Self-attention mechanism apportions varying weights enabling the model to flexibly select important information. The formulation of the self-attention mechanism layer is given by:
\begin{equation}
\begin{split}
Y_{V}^{(l)} &= W_{V_{Att}}^{l}X_{V}^{(l-1)} + b_{V_{Att}}^{l}, \\
Att_{V}^{(l)} &= Softmax(Y_{V}^{(l)}) = \frac{e^{Y_{Vij}^{(l)}}}{\sum_{k=1}^{n}{e^{Y_{Vik}^{(l)}}}}, \\
X_{V}^{(l)} &= Att_{V}^{(l)}X_{V}^{(l-1)}
\end{split}
\end{equation}
where $X_{V}^{(l-1)}$ refers to item embedding outputs from the previous layer, $W_{V_{Att}}^{l}$ and $b_{V_{Att}}^{l}$ are the weight matrix and bias vector for the $l$-th layer respectively, while n indicates the count of columns within the matrix $Y_{V}^{(l)}$. \par 
The core operation of the graph convolution layer is to aggregate characteristics of a node's neighboring nodes, enabling the model to get interactions among potential nodes. The formulation of the graph convolution layer is given by:
\begin{equation}
    X_{V}^{(l+1)} = D_{V}^{-1}A_{V}X_{V}^{(l)}W_{V}^{l}
\end{equation}
where $A_{V}$ denotes the global item graph's adjacency matrix, and $D_{V}$ denotes the degree matrix, defined as $D_{V,i,i} = \sum_{j=1}^{n} A_{V,i,j}$, meaning the $i$-th diagonal element of $D_{V}$ is equal to the sum of the elements in the $i$-th row of $A_{V}$. $X_{V}^{(l)}$ represents $l$-th layer's item embeddings, with $W_{V}^l$ being the parameter matrix for $l$-th layer. \par 
After passing through $L$ layers of attention and graph convolution, the resulting item embeddings are obtained by computing the mean of the item embeddings across all layers:
\begin{equation}
    X_{V} = \frac{1}{L+1} \sum_{l=0}^{L} X_{V}^{(l)}
\end{equation}\par
The session representation is derived by combining the embeddings of items within the session. According to GCE-GNN \cite{wang2020global}, to more effectively capture the order of items throughout a session, a learnable position matrix $P_{r}=[p_{1}, p_{2}, ..., p_{m}]$ is concatenated with item embeddings. Here, $m$ denotes the current session length, and $p_{t}\in\mathbb{R}^d$ is the positional vector associated with the $t$-th position in the session. This helps the model better understand the specific order in the session. In session $s=\{v_{1}^s, v_{2}^s, ..., v_{m}^s\}$, the formulation of the $t$-th item embedding is as follows:
\begin{equation}
    x_{V}^{t*} = \tanh(W_{1}[x_{V}^t \| p_{m-t+1}] + b)
\end{equation}
$W_{1} \in \mathbb{R}^{d \times 2d}$ and $b \in \mathbb{R}^{d}$ represent the adjustable parameters. A soft attention mechanism is commonly used to determine the significance of each item embedding:
\begin{equation}
    a_{t} = q^T \sigma(W_{2} x_{s} + W_{3} x_{V}^{t*} + c), \quad \theta_{V} = \sum_{t=1}^m a_{t} x_{V}^{t*}
\end{equation}
The embedding of the session $s$ is $x_{s}$, which is derived from the mean of the item embeddings within the session, i.e., $x_{s} = \frac{1}{m} \sum_{t=1}^m x_{V}^{t*}$. $q, c \in \mathbb{R}^d$ and $W_{2}, W_{3} \in \mathbb{R}^{d \times d}$ constitute the attention parameters that are employed to determine the weights $a_{t}$ for the items. The session embedding $\theta_{V}$ incorporates the positional information of item embeddings, better expressing the session intent. \par
Drawing from the acquired representations, the recommendation score for each item $v_{i} \in V$ is computed as the dot product of the session $s$ representation with the candidate item $x_{i}$ representation:
\begin{equation}
\hat{z} = {\theta_{V}^{s}}^{T} x_{i}
\end{equation}
Finally, the scores for the items are normalized through the softmax function:
\begin{equation}
\hat{y} = softmax(\hat{z})
\end{equation}\subsection{Single Positive Optimization Loss}
Alignment and uniformity are two important properties for evaluating the quality of representations in contrastive learning. Features with good alignment bring similar users and items closer in the feature space, making it easier to capture their associations and similarities. Features with good uniformity ensure that diverse user and item characteristics have more extensive dispersion in the feature space, which helps to learn more general and discriminative representations. This model adopts a combination of the cross-entropy loss ($L_{ce}$) and the single positive optimization loss ($L_{spl}$):
\begin{equation}
L = L_{ce} + \beta L_{spl}
\end{equation}
where $\beta$ is a hyperparameter that adjusts the significance of the two loss components. The formulation of the cross-entropy loss $L_{ce}$ is given by
\begin{equation}
L_{ce} = -\sum_{i=1}^{n} (y_i \log(\hat{y_{i}}) + (1 - y_i) \log(1 - \hat{y_i}))
\end{equation}
$y_i$ represents one-hot encoded vector corresponding to the true labels. The goal of $L_{ce}$ is to improve the accuracy of predicting the next candidate item in a session by classifying similar samples into the same category during training, which is similar to the optimization objective of alignment in contrastive learning. \par
\begin{figure}
\includegraphics[width=\textwidth]{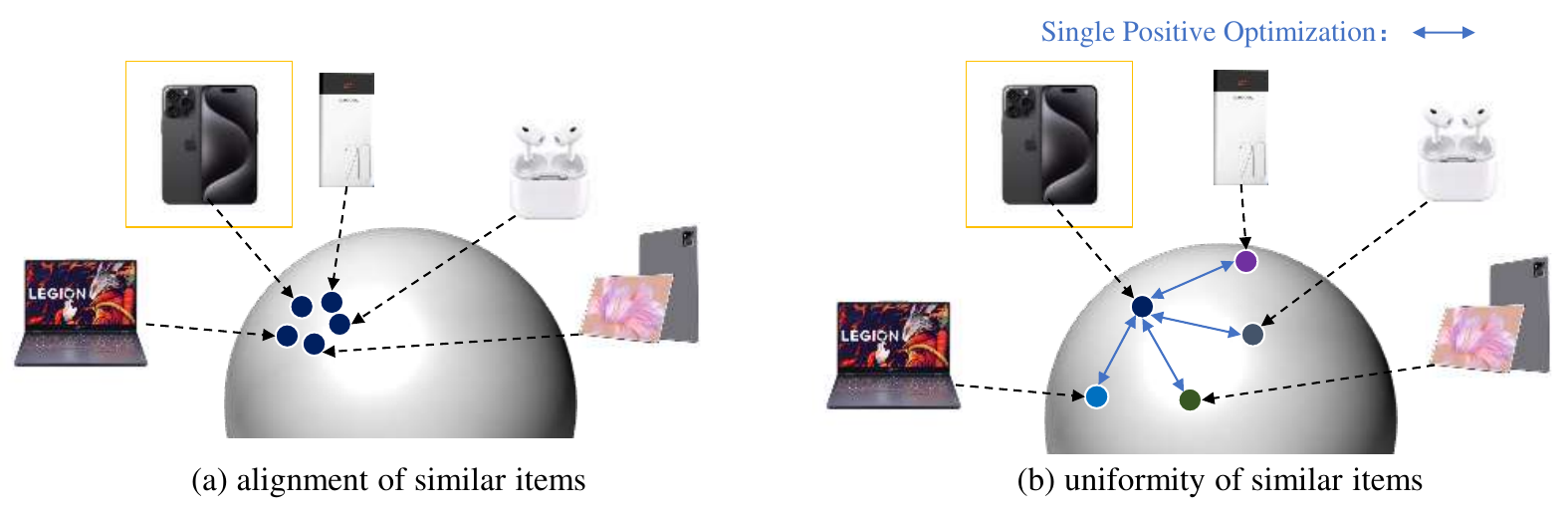}
\caption{An illustration of the principle of single positive optimization. (a) Without single positive optimization, similar items are closely aligned in the feature space. (b) With single positive optimization, the item representation space achieves better uniformity, providing a valuable complement to cross-entropy loss by enhancing the dispersion of item features.}\label{fig3}
\end{figure}

$L_{spl}$ aims to optimize the uniform distribution of item representations in the feature space. By ensuring a certain degree of dispersion among different sample features, $L_{spl}$ enhances the model's capacity for generalization and broadens its utility. According to the self-contrastive learning method \cite{shi2024self}, for a collection of $n$ item representations $X$, the single positive optimization loss is computed with the following objective:
\begin{equation}
L_{spl} = -\sum_{i=1}^{n} \log \frac{g(x_i, x_i)}{\sum_{j=1}^{n} g(x_i, x_j)}
\end{equation}
where the function $g(x_{i}, x_{j})$ is calculated by $\exp(\frac{sim(x_{i}, x_{j})}{\tau})$, where the temperature parameter $\tau$ governs the exponent of the cosine similarity. This loss function $L_{spl}$ employs cosine similarity to distinguish all items across the unit hypersphere. The representation of each item is considered the only positive instance, with the representations of all other items classified as negative instances (as shown in Figure ~\ref{fig3}).
The single positive optimization enhances the even distribution of item representations by adding a self-contrastive loss. This objective directly penalizes item representations that are too close to each other, ensuring that each item representation is well separated in the space. In conclusion, the complete procedure of SPGL is outlined within Algorithm \ref{alg:AOA}.

\begin{algorithm}[!h]

    \caption{The entire workflow of the SPGL model}
    \label{alg:AOA}
    \renewcommand{\algorithmicrequire}{\textbf{Input:}}
    \renewcommand{\algorithmicensure}{\textbf{Output:}}
    \begin{algorithmic}[17]
        \REQUIRE  node embedding vectors $X$, Session data $S$ 
        \ENSURE List of recommended items.    
        
        \STATE  Generate the global item graph $G_{g}$ using the formulas from Eq.(1) to Eq.(3);
        
        \FOR{every iteration}
            \FOR{every batch}
           \STATE Compute item embedding vectors and session representation vectors using the formulas from Eq.(4) to Eq.(8); 
                \FOR{each session s}
            \STATE Compute candidate item scores by following Eq.(9)-Eq.(10);
            \STATE Compute cross-entropy loss and single positive optimization loss using Eq.(12) and Eq.(13) respectively;
                \ENDFOR
            \STATE Optimize the total objective using the formulas in Eq.(11);
            \ENDFOR
        \ENDFOR
        
    \end{algorithmic}
\end{algorithm}

\section{Experiment}
Here, we begin by presenting the datasets, evaluation metrics, and comparative models that were employed in our experimental analysis. Next, we provide details on the parameter settings and overall performance of SPGL model. Ultimately, we evaluate components of SPGL model and examine the impact of hyperparameters on its performance.

\subsection{Datasets}
The performance of our model, SPGL, is evaluated using three widely used benchmark datasets for session-based recommendation: Tmall, RetailRocket, and Diginetica. Tmall dataset consists of user shopping records from the Tmall e-commerce platform. RetailRocket dataset, made available by an e-commerce firm on Kaggle, contains user browsing activities over six months. Diginetica dataset consists of typical transaction data and originates from the CIKM Cup 2016. \par
To facilitate comparison, we align with the experimental setup found in \cite{wu2019session,xia2021selfco}, excluding all sessions that are one-item long and items that have been interacted with fewer than five times. The test set consists of the most recent data (e.g., data from the last week), while earlier data serves as the training set. Then, for each session $s=\{v_{1}^s, v_{2}^s, ..., v_{m}^s\}$, the generated sequence and its label are $([v_{1}^s, v_{2}^s], v_{3}^s)$, ..., $([v_{1}^s, ..., v_{m-1}^s], v_{m}^s)$, where the label for each sequence is determined by the item that was ultimately clicked. Datasets' statistical details are presented in Table \ref{tab1}.
\begin{table}[h]
\vspace{-0.1cm} 
\centering
\caption{Dataset Statistics}\label{tab1}
\setlength{\tabcolsep}{12pt} 
\renewcommand{\arraystretch}{1.1} 
\resizebox{\linewidth}{!}{
\begin{tabular}{c|c|c|c|c}
\hline
\bfseries \textbf{Dataset} & \bfseries \textbf{Train} & \bfseries \textbf{Test} & \bfseries \textbf{Items} & \bfseries \textbf{Mean lengths}\\
\hline
Tmall & 351,268 & 25,898 & 40,728 & 6.69\\
RetailRocket & 433,643 & 15,132 & 36,968 & 5.43\\
Diginetica & 719,470 & 60,858 & 43,097 & 5.12\\
\hline
\end{tabular}}
\vspace{-0.4cm} 
\end{table}

\subsection{Evaluation Metrics and Baselines}
Following \cite{wu2019session,wang2020global,xia2021self}, we employ P@K and MRR@K as metrics to assess the recommendation results, with $K$ being either 10 or 20. \par
To construct a comprehensive comparison, we choose several competitive methods of different categories as baselines: (1) Sequential-based methods, including FPMC \cite{rendle2010factorizing}, GRU4Rec \cite{hidasi2015session}, NARM \cite{li2017neural}, STAMP \cite{liu2018stamp}, and SASRec \cite{kang2018self}; (2) GNN-based methods, including SR-GNN \cite{wu2019session}, GC-SAN \cite{xu2019graph}, CSRM \cite{wang2019collaborative}, GCE-GNN \cite{wang2020global}, MGIR \cite{han2022multi}, SPARE \cite{peintner2023spare} and HICN \cite{sun2024exploiting}; (3) Time-enhanced methods, including TASRec\cite{zhou2021temporal}; (4) Graph contrastive learning methods, including S\textsuperscript{2}-DHCN \cite{xia2021self} and COTREC \cite{xia2021selfco}; (5)
Adaptive loss function methods, including MTAW \cite{ouyang2023mining}.

\subsection{Parameter Settings}
Following prior research, we configure the embedding dimension and batch size to 100. We utilize the Adam optimizer, setting the learning rate to 0.001, and use an $L_{2}$ regularization factor of $10^{-5}$. We set the hop range of the global item graph is to 3. For the Tmall dataset, we use 3 graph convolutional layers, while for RetailRocket and Diginetica, we use 5 layers. In the single positive optimization component, we set the temperature parameter $\tau$ to 0.1 and the single positive optimization loss parameter $\beta$ to 75 for Tmall, 1 for RetailRocket, and 0.75 for Diginetica.

\subsection{Overall Performance}
In Table \ref{tab2}, we present the experimental outcomes for overall performance, highlighting the top results in boldface and the runner-up results with underlines. Results marked  with * denote our own reproductions. Based on the outcomes, we can infer several conclusions. \par
Traditional machine learning methods like FPMC often underperform compared to deep learning techniques because they fail to grasp long-term dependencies. In sequential methods, NARM outperforms GRU4REC by adding an attention mechanism after the RNN to capture user intent. STAMP further improves on NARM by replacing the RNN entirely with attention, giving more focus to relevant items. CSRM leads in performance among these methods by integrating collaborative filtering insights from additional sessions. \par
Graph neural network methods generally outperform sequence-based ones, as they capture spatial structural patterns that sequences may miss. Specifically, GC-SAN, with its self-attention mechanisms, surpasses SR-GNN, and GCE-GNN outperforms GC-SAN by effectively integrating local session details with the global neighbor graph. SPARE and HICN emphasize the importance of graph construction for feature extraction. $S^2$-DHCN and COTREC use contrastive learning with inter-session and intra-session graphs, leveraging self-supervised learning for better performance. MTAW's adaptive weighted loss function also notably enhances recommendation effectiveness. \par
Our proposed SPGL model nearly surpasses all the baseline models across the datasets. Notably, it significantly outperforms other models on the Tmall dataset, demonstrating the efficacy of single positive optimization with graph convolutional networks when applied to real e-commerce data. Relative to the robust benchmark of COTREC, SPGL shows competitive performance and has a simpler structure that does not require constructing a two-branch architecture. Although COTREC attains similar outcomes on the Diginetica dataset, its complex structure leads to reduced transferability. HICN performs slightly  better on RetailRocket, but this model requires complex hypergraph construction. \par
We also introduce a variant called SPGL(CCL), where we replace the single positive optimization with contrastive learning, utilizing the Cross-layer Contrastive Learning (CCL) approach from \cite{yu2023xsimgcl}. SPGL(CCL) uses the initial item embeddings before GNNs and the final item embeddings after GNNs as positive pairs in this cross-layer contrastive setup. Although it has achieved decent results, SPGL(CCL) performs worse compared to SPGL, demonstrating the effectiveness of the single positive optimization. 

\begin{table}[h]
\vspace{-0.1cm} 
\centering
\caption{The performance of all comparison methods across three datasets.}
\label{tab2}
\renewcommand{\arraystretch}{1.1} 
\resizebox{\linewidth}{!}{
\begin{tabular}{l|cccc|cccc|cccc}
\hline
\multirow{2}{*}{\textbf{Method}} & \multicolumn{4}{c|}{\textbf{Tmall}} & \multicolumn{4}{c|}{\textbf{RetailRocket}} & \multicolumn{4}{c}{\textbf{Diginetica}} \\
\cline{2-13}
 & \textbf{P@10} & \textbf{MRR@10} & \textbf{P@20} & \textbf{MRR@20} & \textbf{P@10} & \textbf{MRR@10} & \textbf{P@20} & \textbf{MRR@20} & \textbf{P@10} & \textbf{MRR@10} & \textbf{P@20} & \textbf{MRR@20} \\
\hline
\textbf{FPMC (WWW'10)} & 13.10 & 7.12 & 16.06 & 7.32 & 25.99 & 13.38 & 32.37 & 13.82 & 15.43 & 6.20 & 26.53 & 6.95 \\
\textbf{GRU4Rec (ICLR'16)} & 9.47 & 5.78 & 10.93 & 5.89 & 38.35 & 23.27 & 44.01 & 23.67 & 17.93 & 7.33 & 29.45 & 8.33 \\
\textbf{NARM (CIKM'17)} & 19.17 & 10.42 & 23.30 & 10.70 & 42.07 & 24.88 & 50.22 & 24.59 & 35.44 & 15.13 & 49.70 & 16.17 \\
\textbf{STAMP (KDD'18)} & 22.63 & 13.12 & 26.47 & 13.36 & 42.95 & 24.61 & 50.96 & 25.17 & 33.98 & 14.26 & 45.64 & 14.32 \\
\textbf{SASRec (ICDM'18)} & 21.91 & 11.25 & 27.72 & 12.11 & 37.55 & 22.12 & 45.85 & 23.39 & 35.84 & 14.55 & 48.78 & 17.22 \\
\textbf{SR-GNN (AAAI'19)} & 23.41 & 13.45 & 27.57 & 13.72 & 43.21 & 26.07 & 50.32 & 26.57 & 36.86 & 15.52 & 50.73 & 17.59 \\
\textbf{CSRM (SIGIR'19)} & 25.54 & 13.62 & 29.46 & 13.96 & 43.47 & 25.58 & 51.02 & 26.19 & 36.59 & 15.41 & 50.55 & 16.38 \\
\textbf{GC-SAN (IJCAI'19)} & 24.78 & 13.55 & 28.72 & 13.43 & 43.53 & 26.03 & 50.71 & 25.76 & 37.86 & 16.89 & 50.84 & 17.79 \\
\textbf{GCE-GNN (SIGIR'20)} & 28.01 & 15.08 & 33.42 & 15.42 & 48.22 & 28.36 & 55.78 & 28.72 & \underline{41.16} & 18.15 & 54.22 & 19.04 \\
\textbf{TASRec (SIGIR'21)} & 25.72 & 14.22 & 29.58 & 14.51 & 46.32 & 27.22 & 54.23 & 28.37 & 39.85 & 17.19 & 52.53 & 18.22 \\
\textbf{S\textsuperscript{2}-DHCN (AAAI'21)} & 26.22 & 14.60 & 31.42 & 15.05 & 46.15 & 26.85 & 53.66 & 27.30 & 39.87 & 17.53 & 53.18 & 18.44 \\
\textbf{COTREC (CIKM'21)} & 30.62 & 17.65 & 36.35 & 18.04 & 48.61 & 29.46 & 56.17 & 29.97 & \textbf{41.88} & \underline{18.16} & \underline{54.18} & \underline{19.07} \\
\textbf{MGIR (SIGIR'22)} & 30.71 & 17.03 & 36.31 & 17.42 & 47.90 & 28.68 & 55.35 & 29.20 & 40.63 & 17.86 & 53.73 & 18.77 \\
\textbf{MTAW (SIGIR'23)} & 31.67 & 18.90 & 37.17 & 19.14 & 48.41 & \underline{29.96} & 56.39 & \underline{30.52} & - & - & - & - \\
\textbf{SPARE (RecSys'23)} & \underline{33.61} & \underline{19.78} & \underline{39.28} & \underline{20.07} & 49.07 & 29.75 & 56.91 & 30.22 & 40.70* & 17.85* & 53.99* & 18.78* \\
\textbf{HICN (SDM'24)} & 31.31 & 18.90 & 35.48 & 19.17 & \underline{49.74} & 29.81 & \textbf{57.85} & 30.37 & - & - & - & - \\
\textbf{SPGL(CCL)} & 31.88 & 18.04 & 38.49 & 18.48 & 49.12 & 28.93 & 57.29 & 29.47 &40.72 &17.84 & 54.30 & 18.77 \\
\textbf{SPGL} & \textbf{35.06} & \textbf{20.72} & \textbf{39.64} & \textbf{20.96} & \textbf{49.93} & \textbf{30.23} & \underline{57.40} & \textbf{30.77} & 41.11 & \textbf{18.25} & \textbf{54.33} & \textbf{19.13} \\
\hline
\end{tabular}}
\vspace{-0.9cm} 
\end{table}

\subsection{Ablation Study}
In order to assess the contribution of each component within SPGL model, we conduct comparisons among three different variations: SPGL-SCL, SPGL-Att and SPGL-npos. The SPGL-SCL variant removes the single positive optimization loss function. The SPGL-Att variant removes the self-attention mechanism in the global graph convolution process. The SPGL-npos variant removes the reverse positional information. \par
\begin{figure}[htbp]
    \vspace{-0.1cm} 
    \centering
    \begin{minipage}[t]{0.45\textwidth}
        \centering
        \includegraphics[width=\linewidth]{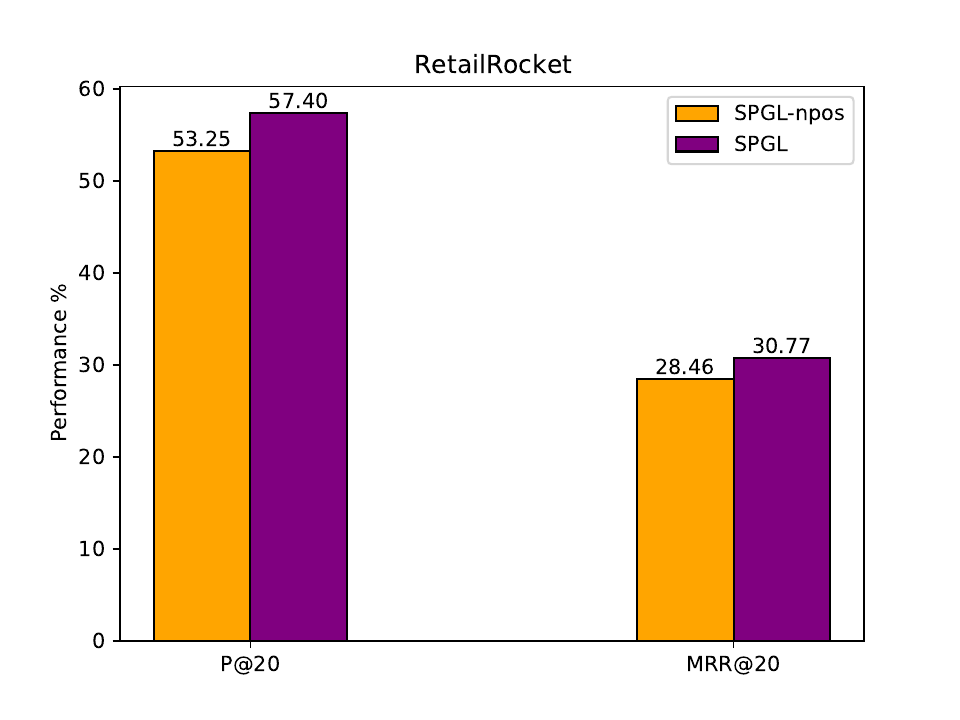}
    \end{minipage}%
    \hfill
    \begin{minipage}[t]{0.45\textwidth}
        \centering
        \includegraphics[width=\linewidth]{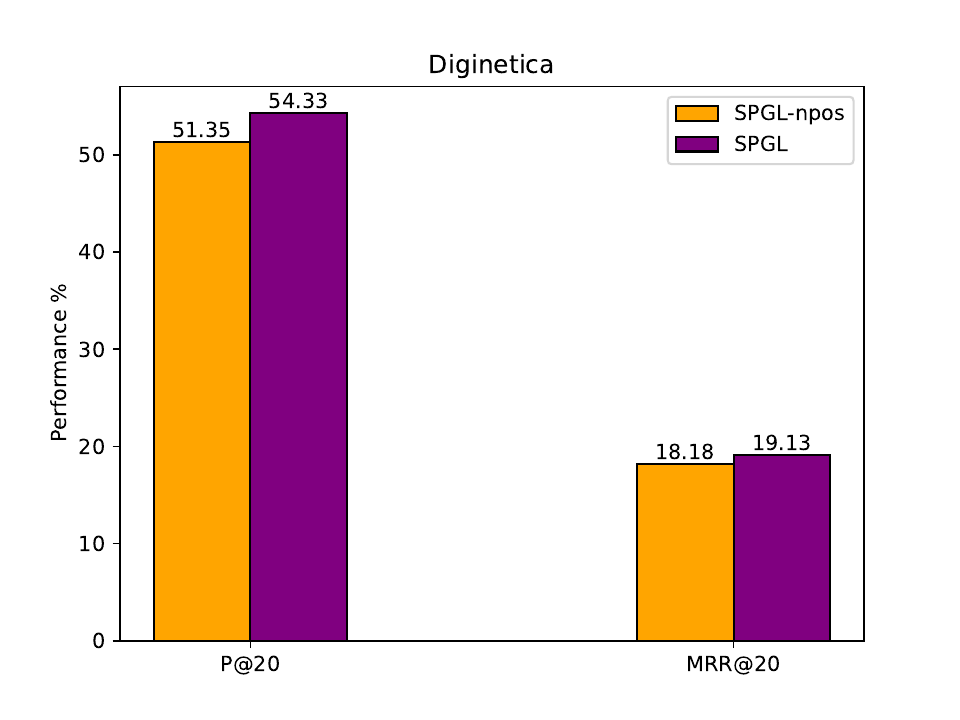}
    \end{minipage}
    \caption{Impact of the reverse positional information}
    \label{fig4}
    \vspace{-0.8cm} 
\end{figure}

According to the results in Table \ref{tab3}, the single positive optimization is identified as the key factor enhancing the model's performance. The self-attention mechanism provides some performance improvement; however, it leads to a decline in the MRR@20 metric on the RetailRocket dataset, likely due to the amplification of noisy data from diverse user behaviors. 
On datasets with shorter session lengths like RetailRocket and Diginetica, the reverse positional information significantly enhances model performance, as shown in Figure \ref{fig4}. This highlights the importance of item temporal order.

\begin{table}[h]
\vspace{-0.1cm} 
\centering
\caption{Performance Comparison of Different SPGL Variants}\label{tab3}
\setlength{\tabcolsep}{12pt} 
\renewcommand{\arraystretch}{1.1} 
\resizebox{\linewidth}{!}{
\begin{tabular}{c|cc|cc|cc}
\hline
\multirow{2}{*}{\textbf{Method}} & \multicolumn{2}{c|}{\textbf{Tmall}} & \multicolumn{2}{c|}{\textbf{RetailRocket}} & \multicolumn{2}{c}{\textbf{Diginetica}} \\
\cline{2-7}
 & \textbf{P@20} & \textbf{MRR@20} & \textbf{P@20} & \textbf{MRR@20} & \textbf{P@20} & \textbf{MRR@20} \\
\hline
SPGL-SCL & 36.46 & 17.77 & 57.33 & 29.62 & 54.24 & 18.74 \\
SPGL-Att & 38.43 & 20.85 & 57.33 & \textbf{30.85} & 54.01 & 19.09 \\
SPGL & \textbf{39.64} & \textbf{20.96} & \textbf{57.40} & 30.77 & \textbf{54.33} & \textbf{19.13} \\
\hline
\end{tabular}}
\vspace{-0.9cm} 
\end{table}

\subsection{Impact of Hyperparameters}
\begin{figure}[htbp]
    \centering
    \begin{minipage}[t]{0.45\textwidth}
        \centering
        \includegraphics[width=\linewidth]{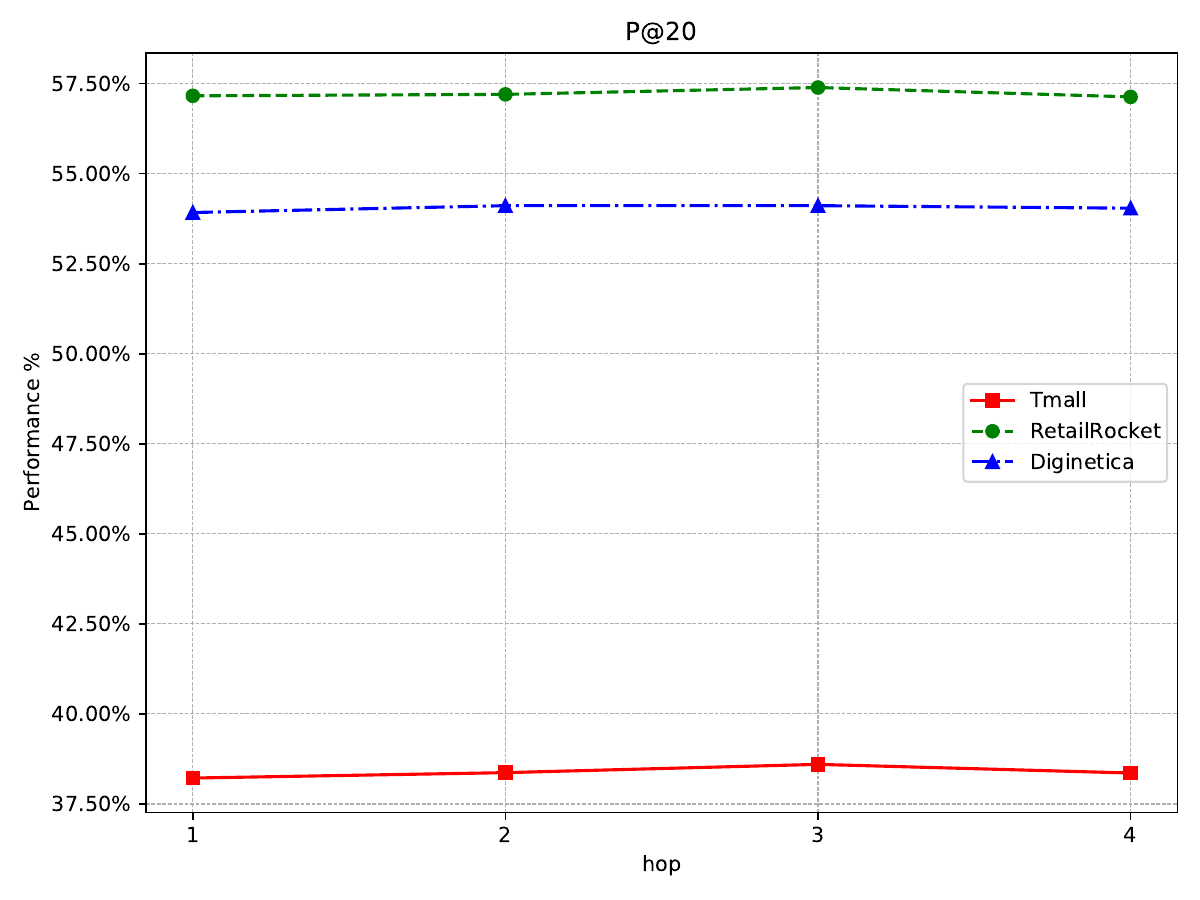}
        \caption{Impact of the hop range}
        \label{fig:hop}
    \end{minipage}%
    \hfill
    \begin{minipage}[t]{0.45\textwidth}
        \centering
        \includegraphics[width=\linewidth]{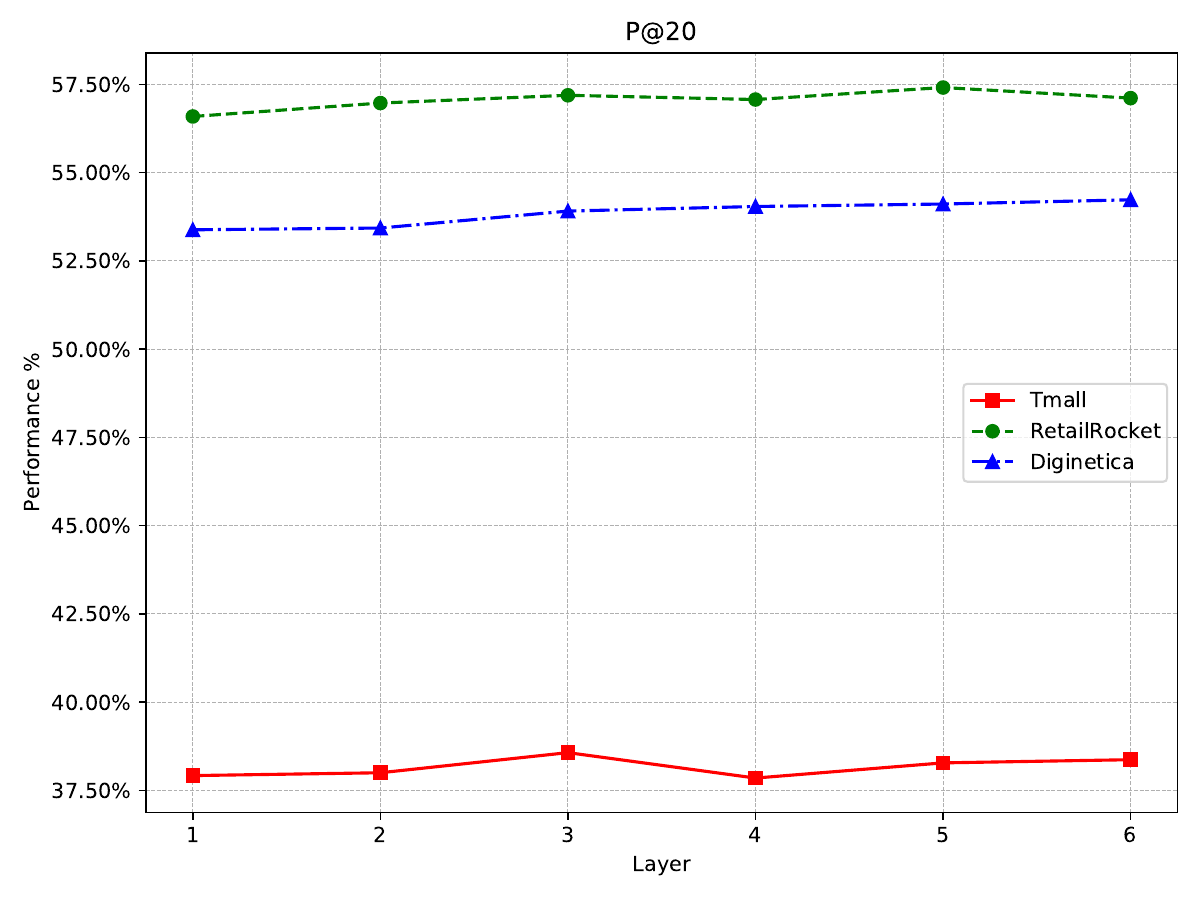}
        \caption{Impact of the number of layers}
        \label{fig:layer}
    \end{minipage}

\end{figure}

The number of hops in the adjacency matrix is pivotal in constructing the directed global item graph. As illustrated in Figure \ref{fig:hop}, the performance is optimal with a hop value of 3 across three datasets. \par Considering the influence of session length on model performance and the experimental outcomes shown in Figure \ref{fig:layer}, we found that using 3 convolutional layers for Tmall and 5 for RetailRocket and Diginetica yields the best results.



\begin{figure}[htbp]
    \centering
    \begin{minipage}[t]{0.45\textwidth}
        \centering
        \includegraphics[width=\linewidth]{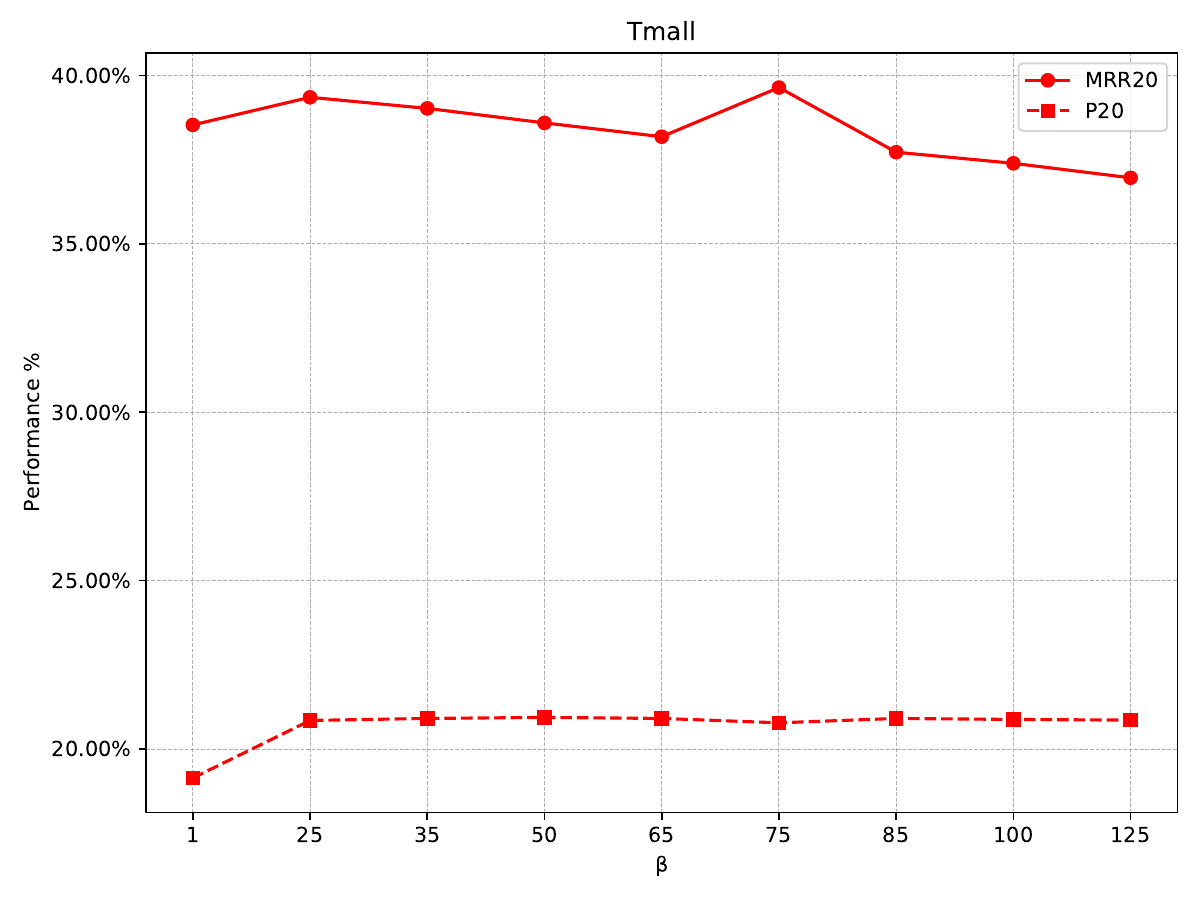}
    \end{minipage}%
    \hfill
    \begin{minipage}[t]{0.45\textwidth}
        \centering
        \includegraphics[width=\linewidth]{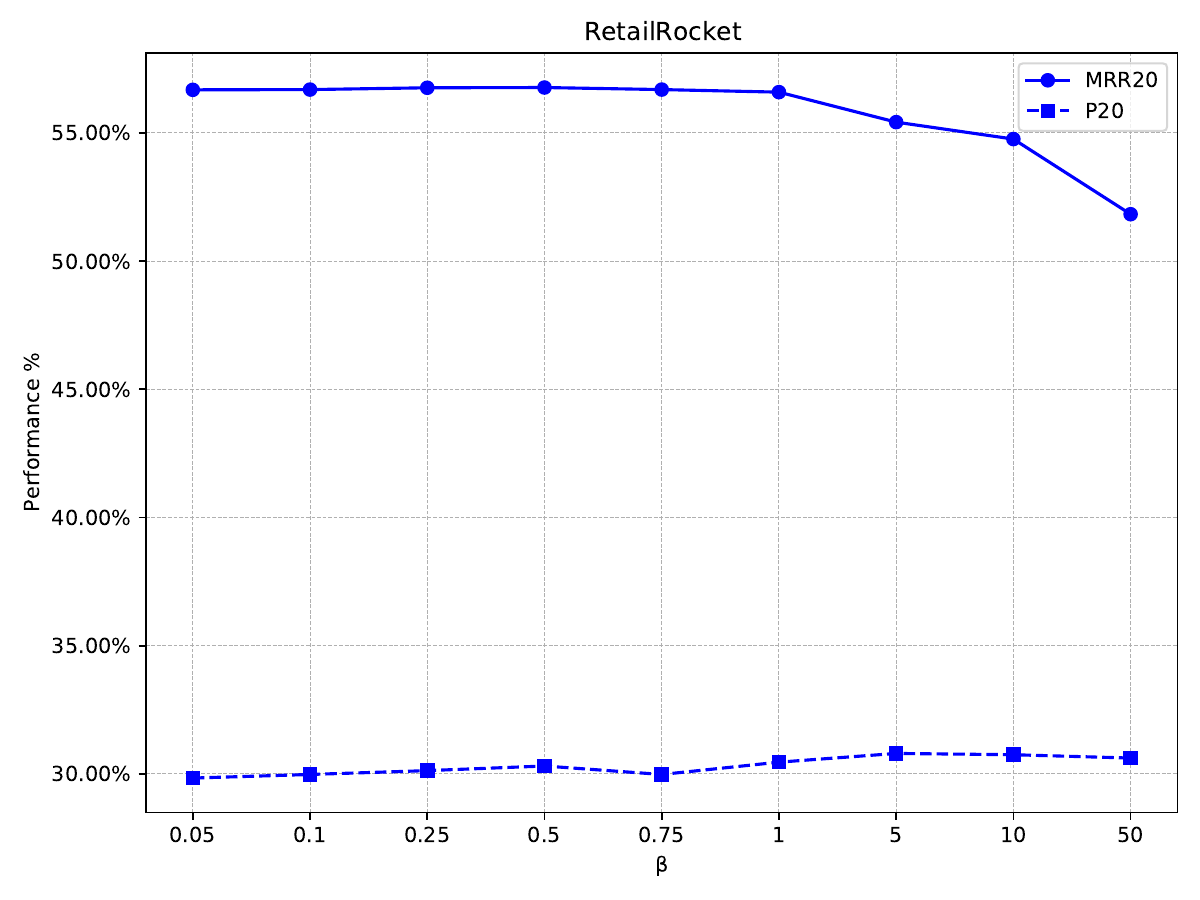}
    \end{minipage}
    \caption{Impact of the single positive optimization loss weight $\beta$}
    \label{fig:beta}
\end{figure}

According to the dataset statistics shown in Table \ref{tab1}, the Tmall dataset has a longer average session length compared to the RetailRocket dataset, which in turn has a longer average session length than the Diginetica dataset. When item counts in the dataset is larger, the constructed collaborative filtering information becomes richer. Intuitively, it is necessary to enhance the uniform distribution of items by setting a larger value for $\beta$. According to the results in Figure \ref{fig:beta}, the best setting is $\beta$=75 on the Tmall dataset, $\beta$=1.0 on the RetailRocket dataset and $\beta$=0.75 on the Diginetica dataset. The $\beta$ parameter balances the uniformity of item distribution and the alignment between items.

\section{Conclusion}
In this paper, we propose a session-based recommendation model that integrates single positive optimization and graph convolutional networks, demonstrating superior performance over existing leading models on three benchmark datasets. The single positive optimization loss aims to enhance the uniformity of item representations in session-based recommendations, supplementing the alignment of similar items achieved by the cross-entropy loss. This approach avoids the complex modeling issues associated with constructing positive and negative samples in contrastive learning, thereby efficiently enhancing session-based recommendation performance through a simple model. Additionally, the self-attention mechanism enables the model to dynamically allocate attention weights in accordance with the data from each node and its neighbors, thereby enhancing the precision of information retrieval. Choosing an appropriate hop count for the adjacency matrix helps to comprehensively capture item information within the spatial domain, while incorporating the reverse positional information helps to model user intent over time. The experimental outcomes demonstrate that the SPGL model is simple and efficient.

\begin{credits}
\subsubsection{\ackname} This research was partially supported by the NSFC (62376180, 62176175, 62302329), the major project of natural science research in Universities of Jiangsu Province (21KJA520004), Suzhou Science and Technology Development Program(SYG202328) and the Project Funded by the Priority Academic Program Development of Jiangsu Higher Education Institutions. \par
This preprint has no post-submission improvements or corrections. The Version of Record of this contribution is published in the Neural Information Processing, ICONIP 2024 Proceedings and is available online at \url{https://doi.org/}.
\end{credits}

\bibliographystyle{splncs04}
\bibliography{3628}

\end{document}